\documentclass[aps,twocolumn,showpacs,superscriptaddress,pra,10pt]{revtex4-1}
\usepackage{graphicx,braket,amsmath,amsfonts,color}
\usepackage[normalem]{ulem}
\usepackage[bookmarks=true,colorlinks=true,breaklinks]{hyperref}
\newcommand*\ham{\hat{H}}
\newcommand*\crt[1]{\hat{a}^\dagger_{#1}}
\newcommand*\dst[1]{\hat{a}^{\phantom{\dagger}}_{#1}}
\newcommand*\vett[1]{{\bf{#1}}}

\newcommand\REVISION[1]{#1}

\begin{document}

\author{Mario Motta}
\email{mmotta@caltech.edu}
\affiliation{Division of Chemistry and Chemical Engineering, California Institute of Technology, Pasadena, CA 91125, USA}

\author{James Shee}
\affiliation{Department of Chemistry, Columbia University, New York, NY 10027, USA}

\author{Shiwei Zhang}
\affiliation{Center for Computational Quantum Physics, Flatiron Institute, New York, NY 10010, USA}
\affiliation{Department of Physics, College of William and Mary, Williamsburg, VA 23187-8795, USA}

\author{Garnet Kin-Lic Chan}
\email{gkc1000@gmail.com}
\affiliation{Division of Chemistry and Chemical Engineering, California Institute of Technology, Pasadena, CA 91125, USA}

\title{Efficient {\em{ab initio}} auxiliary-field quantum Monte Carlo calculations \\
in Gaussian bases via low-rank tensor decomposition}

\begin{abstract}
We describe an algorithm to reduce the cost of  auxiliary-field quantum Monte Carlo (AFQMC) calculations
for the electronic structure problem.
The technique uses a nested low-rank factorization of the electron repulsion integral (ERI).
While the cost of conventional AFQMC calculations in Gaussian bases scales as $\mathcal{O}(N^4)$
where $N$ is the size of the basis, we show that ground-state energies can be computed through
tensor decomposition with \REVISION{reduced memory requirements and sub-quartic} scaling.
The algorithm is applied to hydrogen chains and square grids, water clusters, and hexagonal BN.
\REVISION{In all cases we observe significant memory savings and, for larger systems, reduced, sub-quartic simulation time.}
\end{abstract}

\maketitle

\section{Introduction}

Correlated electronic structure calculations  often require one to store and manipulate tensors, that have
high rank and act on vector spaces of high dimension. Frequently, the input-output and algebraic operations
involving such high-rank tensors constitute a computational bottleneck of the calculations.

The cost of tensor manipulations and storage can be significantly reduced by 
\REVISION{low-rank decompositions 
\cite{Whitten_JCP_1973,Beebe_IJQC12_1977,Yang2011,Hohenstein_JCP_2012,Malone_JCTC_2018}, 
}
in which a higher-rank tensor is represented by contractions of lower-rank tensors.
The most common tensor appearing in Gaussian basis calculations is the rank-$4$ electron-repulsion integral (ERI)
\begin{equation}
  \label{eq:eri}
V_{prqs} = 
\int d{\bf r} d{\bf r}^{\prime} 
\chi_p({\bf r}) \chi_q\left({\bf r}^{\prime}\right) 
\frac{1}{|{\bf r}-{\bf r}^{\prime}|} 
\chi_r({\bf r}) \chi_s\left({\bf r}^{\prime}\right) \, ,
\end{equation}
where the real-valued Gaussian atomic orbitals (AOs) $\{ \chi_p({\bf r}) \}_{p=1}^M$ form a non-orthogonal basis for the one-electron Hilbert space.
Density-fitting (DF) \cite{Whitten_JCP_1973,Dunlap_IJQC_1977,Dunlap_JCP_1979,Werner_JCP_2003}
and modified Cholesky (CD) \cite{Beebe_IJQC12_1977,Koch_JCP118_2003,Aquilante_JCC31_2010} are commonly
applied to obtain a low-rank decomposition of the ERI in the AO basis in terms of a rank-$3$ tensor $L^\gamma_{pr}$,
\begin{equation}
\label{eq:CD}
v_{prqs} \simeq \sum_{\gamma=1}^{N_\gamma} L^\gamma_{pr} L^\gamma_{qs} \quad .
\end{equation}
\REVISION{(To obtain the form (\ref{eq:CD}) in DF, one can apply a Cholesky decomposition or eigenvalue decomposition to the inverse density fitting metric, as is done
in density fitted exchange algorithms~\cite{weigend2002fully,aquilante2007low}).}
Importantly, it is known that the error in such approximations of the ERI decays exponentially with the number of vectors $N_\gamma$, 
and require only $M=\mathcal{O}(N)$ vectors for a fixed error per atom as a function of increasing system size\cite{Sierka_JCP_2003}.
Using the DF or CD approximations reduces the cost of storing the ERI from $\mathcal{O}(N^4)$ to $\mathcal{O}(N^3)$ \cite{Sierka_JCP_2003},
although the computational scaling of most electronic structure methods using DF or CD integrals is not changed.

More recently, several strategies to represent the ERI by  contractions of rank-$2$ tensors have been introduced.
\REVISION{One well known scheme is tensor hyper-contraction, which 
\cite{Hohenstein_JCP_2012,Parrish_JCP_2012,Parrish_PRL_2013,Hohenstein_JCTC_2013,Parrish_JCP_2014,Schumacher_JCTC_2015}
unlike CD or DF, 
can be used to obtain lower-computational scaling in many different electronic structure methods, 
including coupled-cluster \cite{Benedikt_JCP_2011,Shenvi_JCP_2013,Parrish_JCP_2014,Hummel_JCP_2017,Schutski_JCP_2017} 
and Moller-Plesset perturbation theory \cite{Schumacher_JCTC_2015}.
Another recently proposed scheme is nested matrix diagonalization, introduced in Ref.~\cite{Peng_JCTC_2017}.
This has been used to improve quantum computing algorithms for simulating the electronic structure
Hamiltonian \cite{Motta_arxiv_2018}.

In the present work, we explore nested matrix diagonalization 
in the context of the
auxiliary-field quantum Monte Carlo (AFQMC) method in a Gaussian basis \cite{Zhang_PRL90_2003,AlSaidi_JCP124_2006,Motta_WIRES_2018}.
While the cost of conventional Gaussian basis AFQMC scales as $\mathcal{O}(N^4)$ even
after using CD or DF \cite{Purwanto_JCP_2011}, 
we find that low-rank nested matrix diagonalization reduces the \REVISION{computational complexity to sub-quartic (asymptotically cubic), 
while retaining the $\mathcal{O}(N^3)$ storage cost of CD and DF.
As we show, this is because nested matrix diagonalization effectively implements a form of integral screening, by exposing 
it as a low-rank tensor structure.
While cubic scaling is only achievable for very large systems, in the applications presented we always observe a reduction of computation time and sub-quartic scaling.}

It is well known that cubic computational scaling is also achieved in AFQMC calculations}
with plane-waves as the one-electron basis, where the ERI is naturally represented in a factorized form, and the fast Fourier
transform leads to the reduced scaling~\cite{Suewattana_PRB75_2007}.
However, in most scenarios, Gaussian basis sets are more compact than plane-wave bases~\cite{Booth_JCP_2016}.
Thus the current algorithm has the potential to exhibit reduced computational times due to a smaller prefactor 
than plane wave implementations.

The rest of the paper is organized as follows. 
In Section \ref{sec:afqmc} we provide a brief description of the AFQMC method.
In Section \ref{sec:thc}, we describe some of the properties of the nested matrix decomposition and show how a low-rank
approximation can be used to accelerate the most expensive part of an AFQMC simulation, namely, the calculation of the local energy. 
In Section \ref{sec:res}, we assess the performance and accuracy of AFQMC calculations using Gaussian bases
and low-rank decompositions, and conclusions are drawn in Section \ref{sec:conc}.

\section{The AFQMC method}
\label{sec:afqmc}

In this Section, we introduce the AFQMC method and illustrate that the origin of its quartic cost for general electronic structure problems 
lies in the local energy calculation.
Throughout the rest of the paper, we use letters $pqrs$ to indicate a general basis function $\chi_p$ 
(part of an orthogonal or non-orthogonal set over the range  $1\ldots N$), $ijkl$ for particles (indices range from $1\ldots O$), 
$\gamma\mu\nu$ for auxiliary indices associated with the low-rank decompositions (range $1 \ldots M$ for $\gamma$, $1 \ldots \rho_\gamma$ 
for $\mu\nu$). Spin labels are suppressed for compactness.
  
\begin{figure}[ht!]
\centering
\includegraphics[width=0.45\textwidth]{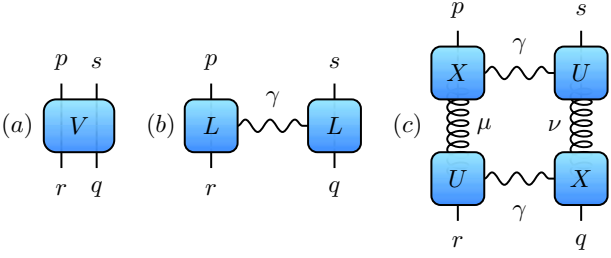}
\caption{(color online) Pictorial illustrations (a) of the rank-$4$ electron repulsion integral (ERI) tensor $V_{prqs}$,
(b) of its Cholesky (CD) or density-fitting (DF) decomposition $V_{prqs} = \sum_{\gamma=1}^{M} L^\gamma_{pr} L^\gamma_{qs}$,
and 
(c) of the low-rank decomposition $V_{prqs} = \sum_{\gamma=1}^{M} \sum_{\mu\nu=1}^{\rho_\gamma} 
X^{\gamma\mu}_p U^{\gamma\mu}_r X^{\gamma\nu}_q U^{\gamma\nu}_s$
used in the present work.
Lines emerging from colored blocks indicate free indices, and lines connecting blocks, indices summed over.
\REVISION{Approximate decompositions in (b, c) break the original ERI into tensors of low rank,
decreasing the memory requirements and cost to evaluate the local energy.}}
\label{fig:lowrank}
\end{figure}

AFQMC \cite{Zhang_PRL90_2003, Motta_WIRES_2018} is a projective quantum Monte Carlo (QMC) method, 
which estimates the ground-state properties of a many-fermion system by statistically sampling the ground-state wavefunction
\begin{equation}
\label{eq:afqmc}
\ket{\Psi_\beta} = 
\frac
{                  e^{-\beta \ham}   \ket{\Phi_T}}
{ \langle \Phi_T | e^{-\beta \ham} | \Phi_T \rangle } 
\overset{ \small{\beta \to \infty} }{ \xrightarrow{\hspace*{0.6cm}} }
\frac{\ket{\Psi_0}}{ \langle \Phi_T | \Psi_0 \rangle} \quad.
\end{equation}
In Eq.~\eqref{eq:afqmc}, $\Psi_0$ is the ground-state wavefunction of the system, 
$\Phi_T$ is an initial wavefunction not orthogonal to $\Psi_0$, which for simplicity we assume to be a single Slater determinant, 
and $\ham$ is the Hamiltonian of the system, 
which without loss of generality \cite{Motta_WIRES_2018} can be written in the form 
\begin{equation}
  \label{eq:h2ndquant}
\hat{H} = E_0 + \sum_{pq} t_{pq} \hat{E}_{pq} + \frac{1}{2} \sum_{prqs} V_{prqs} \hat{E}_{pr} \hat{E}_{qs} \quad.
\end{equation}
$\ham$ is comprised of a constant term, a one-body part written in terms of the 
excitation  operator $\hat{E}_{pq} = \crt{p} \dst{q}$, and a two-body part.
The underlying single-particle basis in Eq.~\eqref{eq:h2ndquant} must be an orthonormal basis.
Thus, when employing a Gaussian AO basis, the AO ERI in Eq.~\eqref{eq:eri} must first be transformed to an orthogonal basis,
as must the DF or CD vectors in the decomposition \eqref{eq:CD}. Using the transformed CD vectors, the two-body part
can be written as a sum of squares of one-body operators,
\begin{equation}
\sum_{prqs} V_{prqs} \hat{E}_{pr} \hat{E}_{qs} = \sum_{\gamma} \hat{v}_\gamma^2
\quad,\quad
\hat{v}_\gamma = \sum_{pr} L_{pr}^\gamma \hat{E}_{pr} 
\quad.
\end{equation}
These are illustrated in Figure \ref{fig:lowrank}(a) and (b).
For sufficiently large $\beta$, expectation values computed over $\Psi_\beta$ yield ground-state averages.
AFQMC projects $\Psi_T$ towards $\Psi_0$ iteratively, writing
\begin{equation}
e^{-\beta \ham} = \left(e^{-\Delta\tau \ham} \right)^n \quad ,
\end{equation}
where $\Delta\tau = \frac{\beta}{n}$ is a small imaginary-time step. 
The propagator is represented, 
through a Hubbard-Stratonovich transformation \cite{Stratonovich_SPD2_1958,Hubbard_PRL3_1959}, as
\begin{equation}
\label{eq:hs}
e^{-\Delta\tau \ham} = \int d{\bf{x}} \, p({\bf{x}}) \, \hat{B}({\bf{x}}) \quad,
\end{equation}
where
\begin{equation}
\hat{B}({\bf{x}}) 
=
\exp\left(-\Delta\tau\, \hat{H}_1 
+ i 
\sqrt{\Delta\tau}\sum_{\gamma=1}^{N_\gamma} x_{\gamma} \hat{v}_{\gamma}\right)
\end{equation}
is an independent-particle propagator that depends on the vector of fields ${\bf{x}}$, 
$p({\bf{x}})$ is the standard normal $M$-dimensional probability distribution 
and $\hat{H}_1 = \sum_{pq} t_{pq} \hat{E}_{pq}$ is the one-body part of $\hat{H}$.
The representation \eqref{eq:hs} maps the original interacting many-fermion system 
onto an ensemble of non-interacting systems subject to a fluctuating potential.
The imaginary-time projection can be realized as an open-ended random walk over paths of auxiliary-fields ${\bf{x}}$
\cite{Zhang_PRL90_2003}.
Importance sampling the trajectories of the random walk leads to a representation of ${\Psi_\beta}$ 
as a stochastic weighted average of Slater determinants,
\begin{equation}
\ket{\Psi_\beta} \simeq \frac{1}{\sum_{w} W_{w}} \sum_{w} W_{w} \frac{ \ket{\Phi_{w}} }{ \langle \Phi_T |\Phi_{w} \rangle }
\quad .
\end{equation}
Because the phase in $\hat{v}_\gamma$ can be complex for general two-body interactions,
AFQMC suffers from a phase problem. This can be controlled using a trial state $|\Phi_T\rangle$ and imposing
the phaseless approximation (Ph) and a real local energy estimator~\cite{Zhang_PRL90_2003,Motta_WIRES_2018};
the error of these approximations vanishes if the trial state is exact.

The accuracy of Ph-AFQMC calculations of ground- and excited-state energies has been extensively benchmarked 
both in \emph{ab initio} studies~\cite{AlSaidi2007,Purwanto_PRB80_2009,Virgus2014,Motta_PRX_2017} 
and lattice models of correlated electrons \cite{LeBlanc_PRX5_2015,Qin2016}.
The random walks take place in the over-complete manifold of Slater determinants, 
in which fermion antisymmetry is maintained by construction in each walker. 
Recently, the Ph-AFQMC has also been extended to the calculation of general ground-state properties, 
energy differences and interatomic forces in realistic materials \cite{Motta_JCTC_2017,Motta_JCP_2017,Shee_JCTC13_2017}.

In \emph{ab initio\/} computations,  the electron repulsion integrals entering into the AFQMC calculation can be obtained in different computational bases, such as  plane-waves and pseudo-potentials 
\cite{Zhang_PRL90_2003,Ma_PRB_2016} or 
 Gaussian type orbitals \cite{AlSaidi_JCP124_2006}.
 This choice of representation is important because it affects the cost of the AFQMC algorithm.
  When plane-waves are used, the standard AFQMC methodology is known to scale as $\mathcal{\tilde{O}}(N^3)$,
  \footnote{Here, and the remainder of the present work, 
we rely on the soft-$\mathcal{O}$ notation, well-established in complexity theory: 
$g(x) = \tilde{\mathcal{O}}\left(f(x)\right)$ if there exists an integer $k$ such that $g(x) = \mathcal{O}\left(f(x) \log^k(x)\right)$.}
  as documented in \REVISION{Appendix~\ref{sec:pw}}. When using a Gaussian basis, on the other hand, state-of-the-art calculations feature $\mathcal{O}(N^4)$ cost.
  The computational bottleneck in both cases tends to be the local energy calculation, which we describe below.

\subsection{Local energy calculation}
\label{sec:thc}

AFQMC calculations require the computation of the following local energy functional for each sample,
\begin{equation}
\label{eq:localenergy}
\mathcal{E}_{loc}(\Phi)=\frac{\langle\Phi_{T}|\ham| \Phi\rangle}{\langle\Phi_{T}| \Phi\rangle}
\quad,
\end{equation}
from which the total energy is obtained as \REVISION{$E=\sum_w W_w E_{loc}(\Phi_w)$}. 
The local energy is also
needed to determine the weights in Ph calculations \REVISION{\cite{Zhang_PRL90_2003,Motta_WIRES_2018,Purwanto_PRB80_2009}.}
The most demanding part of its calculation comes
from the two-body term $\ham_2$ which, from the generalized Wick's theorem 
\cite{Balian_NC64_1969}, can be written as 
\begin{equation}
\label{eq:localenergy2}
\begin{split}
&2 \mathcal{E}_{loc,2}(\Phi) 
=
2 \frac{ \langle \Phi_T | \hat{H}_2 | \Phi \rangle }{ \langle \Phi_T | \Phi \rangle }
= \\
&= \sum_{prqs} V_{prqs} \left(G_{pr} G_{qs}  -
G_{ps} G_{qr} \right)
\,
\end{split}
\end{equation}
where the one-body reduced density matrix (RDM1)
\begin{equation}
\label{eq:rdm}
\begin{split}
G_{pr} 
&= 
\frac{ \langle \Phi_T | \crt{p} \dst{r} | \Phi \rangle }{ \langle \Phi_T | \Phi \rangle }
=
\Big[ \Phi \big( \Phi_T \Phi \big)^{-1} \Phi_T \Big]_{rp} \\
&= 
\sum_{i}\Theta_{ri} {\Phi_T}_{ip}
\\
\end{split}
\end{equation}
is defined in terms of the matrices $\Phi$ (of dimension $N \times O$) and $\Phi_T$ 
($O \times N$) parametrizing 
the Slater determinant and trial wave-function respectively,
\begin{equation}
\begin{split}
| \Phi \rangle &=  \prod_{i} \crt{\phi_i}|\emptyset \rangle 
\quad,\quad 
|\phi_i \rangle = \sum_{p} \Phi_{pi} |\chi_p \rangle \\
\langle \Phi_T | &= \langle \emptyset |  \prod_{i} \dst{\phi^T_i} 
\quad,\quad 
\langle \phi^T_i | = \sum_{p} {\Phi_T}_{ip} \langle \chi_p |  \quad .
\end{split}
\end{equation}
\REVISION{In Eq. \eqref{eq:rdm}, $\Theta = \Phi \big( \Phi_T \Phi \big)^{-1}.$}
Note that the expression Eq.~\eqref{eq:rdm} for the RDM1 sample resembles the expression for the RDM1 of the trial Slater determinant, \REVISION{$G_{T} = \Phi^\dagger_T \Phi_T$},
with one $\Phi_T$ matrix (walker independent) replaced by $\Theta$ (dependent on the walker). Explicit evaluation
of \eqref{eq:rdm} costs $\mathcal{O}(ON^2)$ per sample while the summation in the two-body local energy costs $\mathcal{O}(N^4)$ per sample.
For $N\gg O$, it is more efficient \cite{AlSaidi_JCP124_2006} to first contract the two-body matrix elements with $\Phi_T$,
\begin{equation}
  \bar{V}_{piqj} = \sum_{rs} {\Phi_T}_{ir} {\Phi_T}_{js} V_{prqs},
\end{equation}
which may be carried out once and stored at the start of the AFQMC calculation at a cost of $\mathcal{O}(ON^4+O^2N^3)$. The
local energy then follows as the sum
\begin{equation}
  \label{eq:convafqmc}
2 \mathcal{E}_{loc,2}(\Phi) = \sum_{piqj} \bar{V}_{piqj} \Theta_{pi} \Theta_{qj}
  \end{equation}
at a cost of $\mathcal{O}(O^2N^2)$ per sample. When memory is not a limitation, this is the
most efficient conventional algorithm for local energy evaluation and is the one we compare against in this work.

\begin{figure}[ht!]
\centering
\includegraphics[width=0.95\columnwidth]{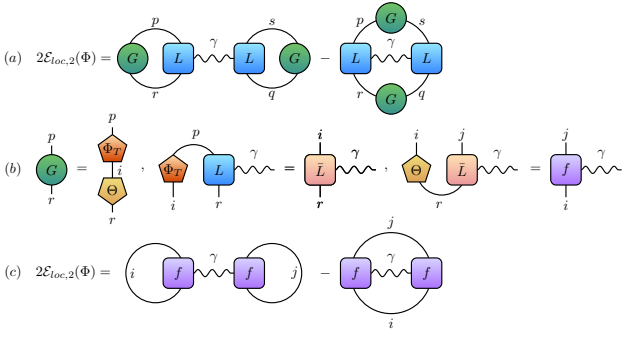}
\caption{(color online) (a) Pictorial representation of the local energy calculation based on the CD decomposition
of the ERI. (b) separable structure of the RDM1, as used in precomputing the tensors $\bar{L}$ and $f$.
(c) Expression of the local energy based on the $f$ tensor.
}
\label{fig:eloc1}
\end{figure}

As mentioned in the introduction, the Cholesky decomposition \eqref{eq:CD}
allows one to significantly reduce the storage requirements by replacing
the 4-index integrals by a truncated set of 3-index quantities. 
However, it does not reduce the computational cost of local energy evaluation.
Inserting \eqref{eq:rdm} into \eqref{eq:localenergy2} and using the CD form in \eqref{eq:CD} (after transformation
to an orthogonal basis) gives
\begin{equation}
  \label{eq:energy_from_f}
2 \mathcal{E}_{loc,2}(\Phi) = \sum_{ij\gamma} f_{ii}^\gamma f_{jj}^\gamma - f_{ij}^\gamma f_{ij}^\gamma,
\end{equation}
with the intermediate $f^\gamma_{ij}$  defined as
\begin{equation}
  \label{eq:ftensor}
  f^\gamma_{ij} = \sum_{pr} \left({\Phi_T}_{ip} L^\gamma_{pr}\right) \Theta_{rj}.
  \end{equation}
This is computed most efficiently by precomputing and storing the quantity in brackets, 
$\bar{L}^{\gamma}_{ir} = \sum_{p} {\Phi_{T}}_{ip} L^\gamma_{pr}$,
at the beginning of the AFQMC run, at cost $\mathcal{O}(ON^2M)$, and subsequently carrying out the second contraction
for each sample with $\mathcal{O}(O^2NM) \sim \mathcal{O}(N^4)$ cost.
However, as $M > N$, the reduced memory cost afforded by CD is offset by an increased 
computational cost of the local energy evaluation, compared with the conventional 
algorithm in \eqref{eq:localenergy2}.
The operations described so far are illustrated diagrammatically in Figure \ref{fig:eloc1}.

To overcome this increased cost, we now describe how we can exploit additional structure in the Cholesky vector $L^\gamma_{pr}$.

\section{Low-rank factorization via nested matrix diagonalization and acceleration of local energy evaluation}
\label{sec:thc}

\REVISION{
Ref. \cite{Peng_JCTC_2017} introduced a truncated nested matrix diagonalization, corresponding to an additional truncated
factorization of $L^\gamma$. This starts from a truncated CD of the ERI, Eq. \eqref{eq:CD},
such that all elements of the residual
\begin{equation}
R_{prqs} = V_{prqs} - \sum_{\gamma} L^\gamma_{pr} L^\gamma_{qs}
\end{equation}
are kept smaller in absolute value than a predefined threshold $\varepsilon_{CD}$.
Note that, since $|R_{prqs}| \leq R_{prpr}$ \cite{Beebe_IJQC12_1977}, bounding $R$
requires computing and bounding its diagonal only.

Then, after} transformation to an orthogonal basis, we carry out
an eigenvalue decomposition of the matrix $L_{pr}^\gamma$ for each $\gamma$,
\begin{equation}
L^\gamma_{pr} = \sum_{\mu} U_{p\mu}^{\gamma} \sigma^\gamma_\mu U_{r\mu}^\gamma \quad ,
\end{equation}
and only eigenvalues larger in absolute value than a predefined threshold $\varepsilon_{ET}$ are kept,
$|\sigma^\gamma_\mu| \geq \varepsilon_{ET}$.
This additional eigenvalue truncation (ET) leads to the approximation
\begin{equation}
\label{eq:thc}
\begin{split}
V_{prqs} &\simeq \sum_\gamma \sum_{\mu\nu} 
(U_{p\mu}^{\gamma} \sigma^\gamma_\mu) U_{r\mu}^\gamma
(U_{q\nu}^{\gamma} \sigma^\gamma_\nu) U_{s\nu}^\gamma \\
&=  \sum_\gamma \sum_{\mu\nu}  
X_{p\mu}^{\gamma} U_{r\mu}^\gamma
X_{q\nu}^{\gamma} U_{s\nu}^\gamma
\end{split}
\end{equation}
where $\rho_\gamma \leq N$ is the number of retained eigenvalues for the matrix $L^\gamma$
and $X_{p\mu}^{\gamma} = U_{p\mu}^{\gamma} \sigma^\gamma_\mu$.
The decomposition \eqref{eq:thc} is diagrammatically illustrated in Figure \ref{fig:lowrank}(c).

\begin{figure*}[ht!]
\centering
\includegraphics[width=0.49\textwidth]{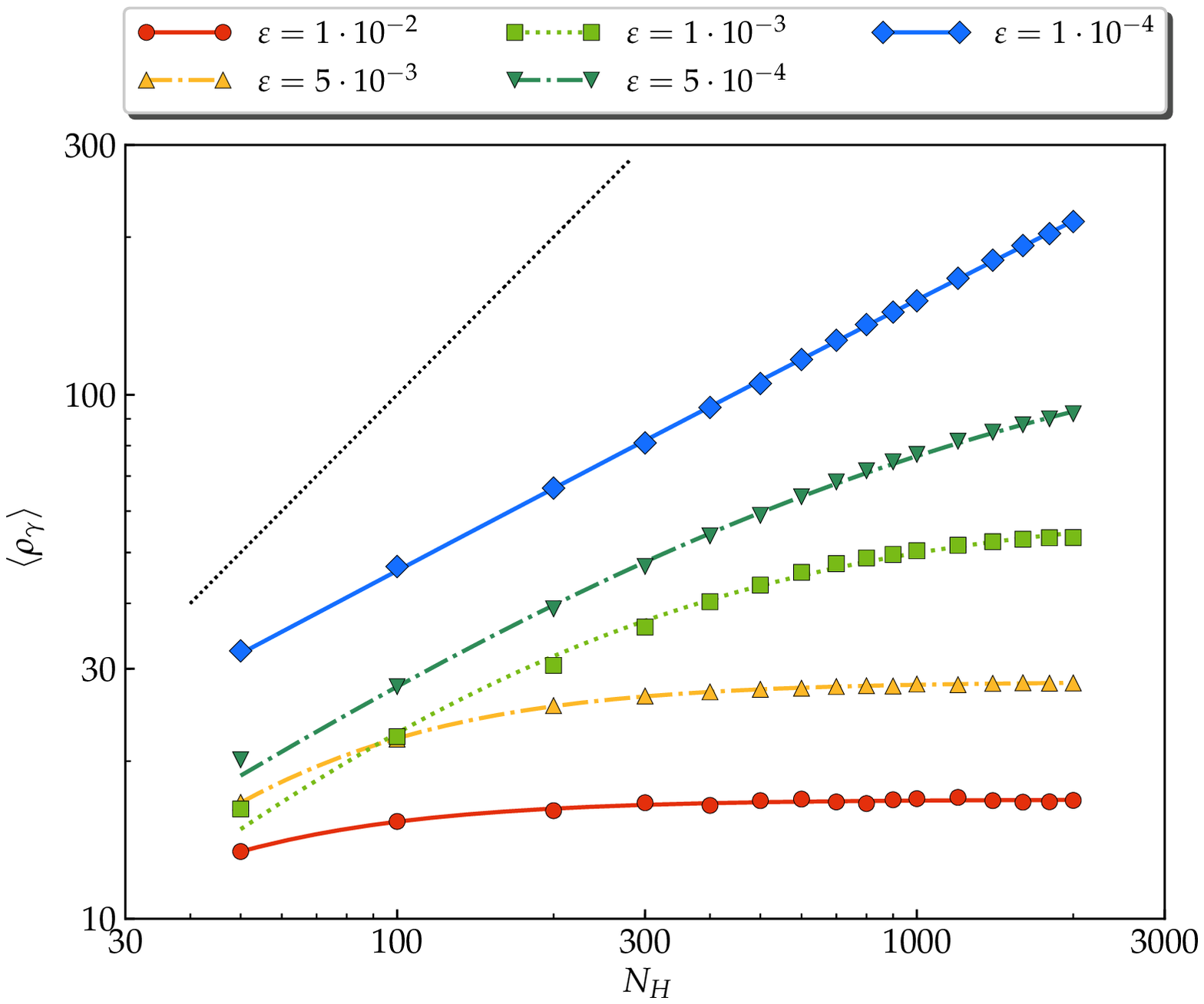}
\caption{(color online) \REVISION{Average number $\langle \rho_\gamma \rangle$
of eigenvalues for H chains on a log-log scale, at the representative bondlength $R=1.8 \mathrm{a_B}$, 
at the STO-6G level, using thresholds $\varepsilon$ between $10^{-4}$ and $10^{-2}$ au.
The black dotted line represents the number $N$ of basis functions, providing an upper bound
for $\langle \rho_\gamma \rangle$. Coloured lines are the result of a fit to $\frac{x^\alpha}
{\beta + \gamma x^\alpha}$. Sub-linear growth is visible in all cases, and saturation is reached
for the looser thresholds.}
}
\label{fig:ng}
\end{figure*}

\REVISION{
In Ref.~\cite{Peng_JCTC_2017} it was suggested, without detailed analysis, that
the average number of eigenvalues of the Cholesky vectors 
\begin{equation}
\langle     \rho_\gamma \rangle = \frac{1}{M} \sum_{\gamma}    \rho_\gamma
\quad,
\end{equation}
grows logarithmically with increasing system size \cite{Peng_JCTC_2017}. In fact, we have found that the
data presented in Ref.~\cite{Peng_JCTC_2017} can be fit just as well by a variety of functional forms, including
by $\alpha N^\beta$  with $\beta \sim 1/2$.
However, we now argue that in large systems, $\langle \rho_\gamma\rangle \to \tilde{\mathcal{O}}(1)$, because the number
of eigenvalues above a given threshold is related to $\frac{1}{N}$ times the number of Coulomb integrals above
an integral threshold, which is asymptotically $\mathcal{O}(1)$. To see this, assume the Gaussian basis
has minimum exponent $\alpha$. Then on length scales longer than $\alpha^{-1/2}$, we can simplify $V_{prqs}$ to 
a two-index quantity
\begin{equation}
\label{eq:moderi}
V_{PQ} = 
\left\{
\begin{array}{ll}
c |P-Q|^{-1} & P \neq Q \\
V_0 & P = Q \\
\end{array}
\right.
\end{equation}
where $V_0$ is related to the maximum exponent of the basis.
The Cholesky  decomposition $V_{PQ}=\sum_\gamma L_P^\gamma L_Q^\gamma$
yields Cholesky vectors that are already ``diagonal'' for each $\gamma$. Thus eigenvalue truncation in Eq.~\eqref{eq:thc}
truncates elements of the Cholesky vectors by absolute value, and the average number of significant eigenvalues $\rho_\gamma$
is the average number of significant elements of the Cholesky vectors $L_P^\gamma$.

\begin{figure*}[t!]
\centering
\includegraphics[width=0.79\textwidth]{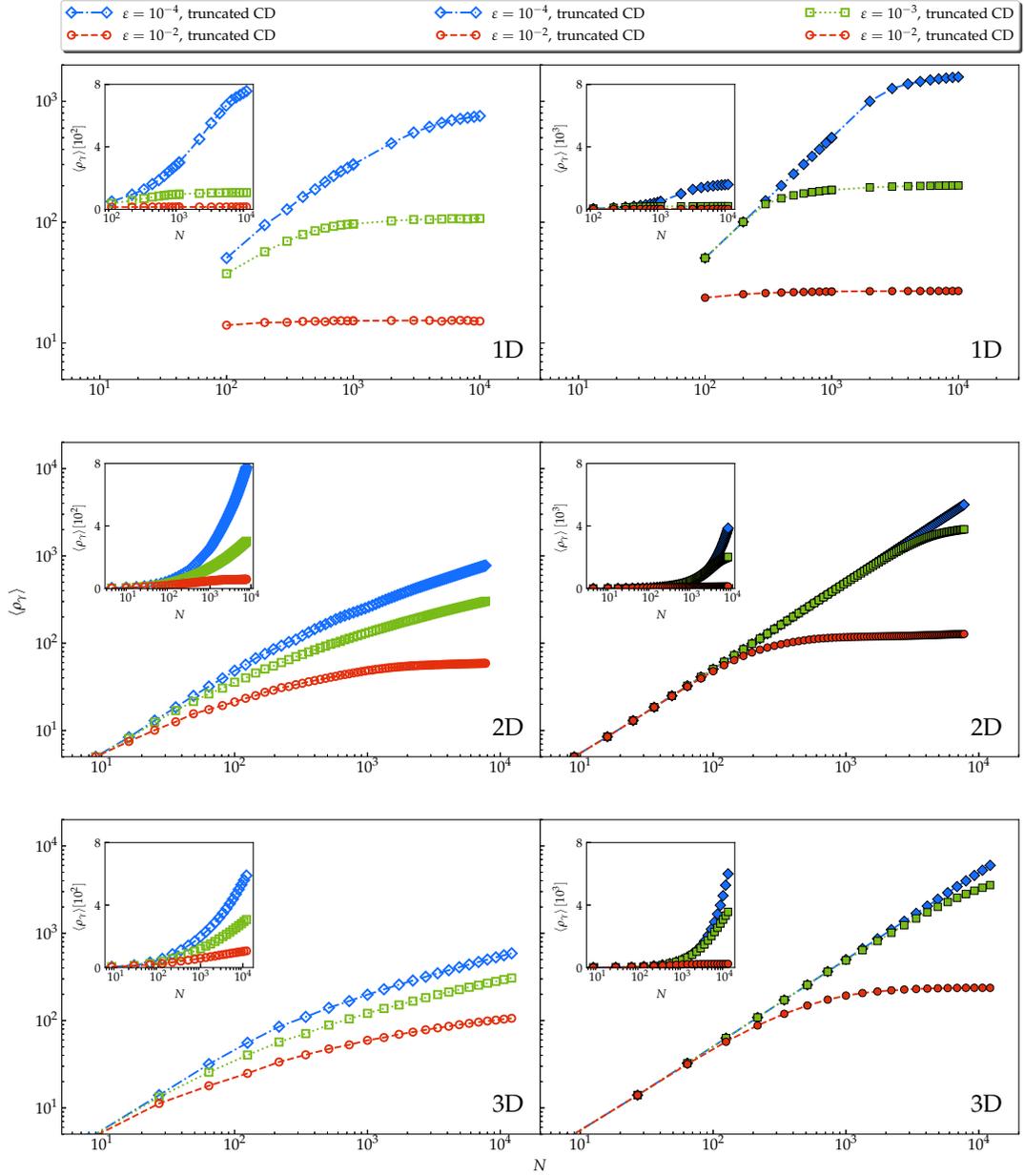}
\caption{
\REVISION{Main plots: Growth of the average number of eigenvalues $\langle \rho_\gamma \rangle$, on a
log-log scale, for the model electron repulsion integral \eqref{eq:moderi} in 1D, 2D, 3D (top to bottom), using 
$\varepsilon = 10^{-2}, 10^{-3}, 10^{-4}$ (red, green, blue). The parameters in \eqref{eq:moderi} 
are $c,V_0=1/2,1$.
We either perform a truncated CD over the exact ERI (left, empty symbols) or an untruncated
CD over the truncated ERI (right, filled symbols). Note that on the log-log scale, the slope $\alpha$ gives 
 $\langle \rho_\gamma\rangle \sim N^\alpha$; in all cases $\alpha < 1$ and asymptotically approaches 0.
Insets: the same quantities, on a log-linear scale. 
}}
\label{fig:modellino1}
\end{figure*}

\begin{figure*}[t!]
\centering
\includegraphics[width=0.79\textwidth]{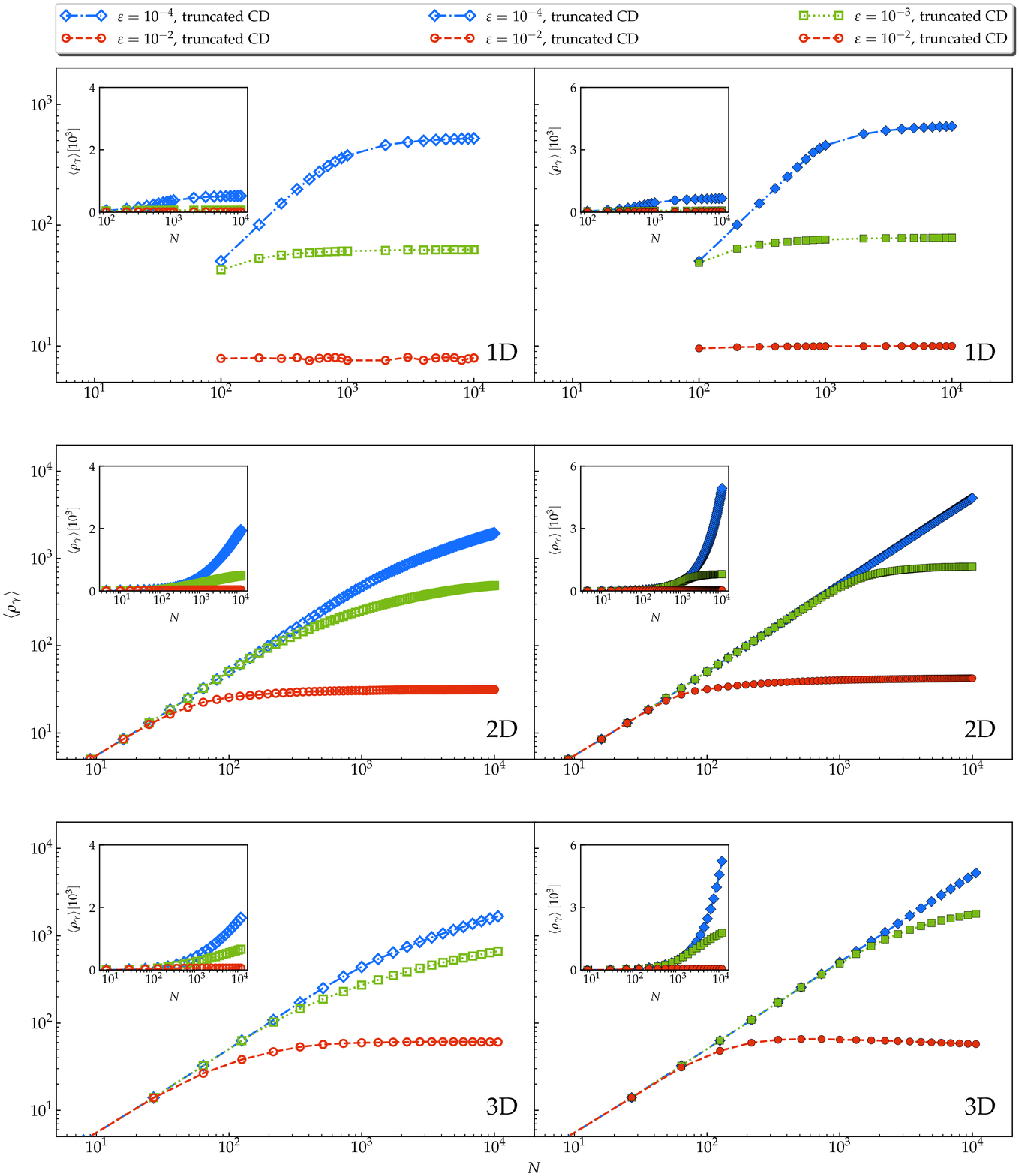}
\caption{
\REVISION{Same as figure \ref{fig:modellino1} but with parameters $c,V_0=1/10,1$.
}}
\label{fig:modellino2}
\end{figure*}

We now introduce a simple model to understand the behaviour of $\rho_\gamma$.
First consider a one-dimensional lattice of Gaussian functions (e.g. $1s$ functions) evenly spaced for simplicity. If 
integral screening is used with threshold $\varepsilon$, we truncate $V_{PQ}$ such that it is a banded matrix of width $w(\varepsilon)$. Then, the
Cholesky vectors are also strictly banded, i.e. $L_p^\gamma = 0, |\gamma-p| > w$, and we rigorously obtain $\rho_\gamma = \mathcal{O}(1)$.
This is not precisely a statement about truncating Cholesky elements of the full (untruncated) Coulomb matrix, but as we see in Fig.~\ref{fig:modellino1} and Fig.~\ref{fig:modellino2} the behaviour of $\langle \rho_\gamma\rangle$ in these two settings is exactly the same. For $V_{PQ}$
corresponding to a general graph, the number of non-zeros of the Cholesky vectors above a threshold is well-studied
as the problem of fill-in generated by threshold-based incomplete Cholesky factorization~\cite{lin1999incomplete}. While rigorous
bounds are difficult to prove, we numerically compute $\langle \rho_\gamma\rangle$ for 2D and 3D cubic lattices both for $V_{PQ}$
first truncated by a threshold $\varepsilon$, as well as for the untruncated $V_{PQ}$. The behaviour is very similar in both cases.
While we cannot rule out nested logarithmic factors such as $\log(N)$ or $\log(\log(N))$,
this numerical evidence strongly suggests that $\langle \rho_\gamma\rangle$ saturates at $\tilde{\mathcal{O}}(1)$, just as it does in 1D.
Finally, the same numerical behaviour can be seen when decomposing
the 4-index integral tensor, which we show in Figure \ref{fig:ng} for hydrogen chains, where we can
reach sufficiently large sizes to see saturation unambiguously for sufficiently large thresholds.
For a fixed truncation accuracy and up to possible logarithmic factors, we consider the evidence
to be strong that $\langle \rho_\gamma\rangle$ saturates to become independent of system size.
}

\subsection{Accelerated local energy evaluation}

The low-rank structure revealed in the Cholesky vectors directly reduces the computational and memory
costs of the AFQMC algorithm. In the present work we choose $\varepsilon_{CD} = \varepsilon_{ET}$,
although the two thresholds can in principle be chosen separately \cite{Peng_JCTC_2017}.

\begin{figure}[ht!]
\centering
\includegraphics[width=0.95\columnwidth]{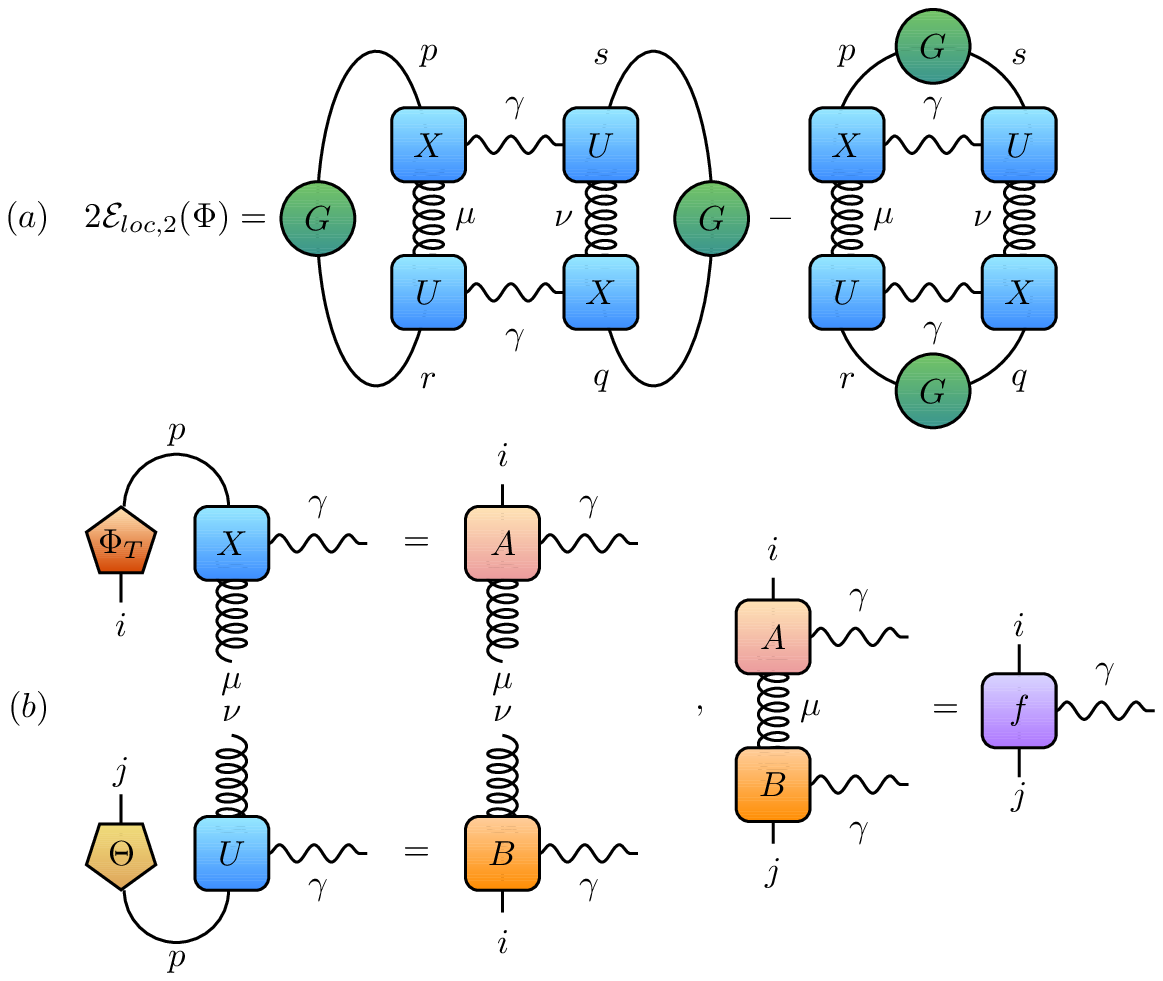}
\caption{(color online) Pictorial illustrations 
(a) of the local energy calculation based on the CD+ET decomposition of the ERI,
(b) of the precomputed and intermediate tensors involved in the calculation.
The final expression for the local energy coincides with the one in Figure \ref{fig:eloc1}(c).
}
\label{fig:elocthc}
\end{figure}

In the case of the local energy, the intermediate $f_{ij}^\gamma$ can be built as (see also Figure \ref{fig:elocthc})
\begin{align}
  \label{eq:ftensor}
  f^\gamma_{ij} &= \sum_{pr\mu} ({\Phi_T}_{ip}
  U_{p\mu}^{\gamma} \sigma^\gamma_\mu) (U_{r\mu}^\gamma  \Theta_{rj}) \nonumber \\
  &= \sum_{\mu} A^{\gamma\mu}_i B^{\gamma\mu}_j
  \end{align}
where $A$ can be evaluated at the beginning of the AFQMC run with cost $\mathcal{O}(NM 
\REVISION{\langle \rho_\gamma \rangle})$,
$B$ is evaluated for each sample with cost 
$\mathcal{O}(ONM \REVISION{\langle \rho_\gamma \rangle})$, and the assembly into $f^\gamma_{ij}$
is $\mathcal{O}(O^2M \REVISION{\langle \rho_\gamma \rangle})$ per sample. \REVISION{For a sublinear
(constant) $\langle \rho_\gamma \rangle$, this is then gives sub-quartic (cubic) cost for the energy evaluation.}

\begin{figure}[h!]
\centering
\includegraphics[width=0.45\textwidth]{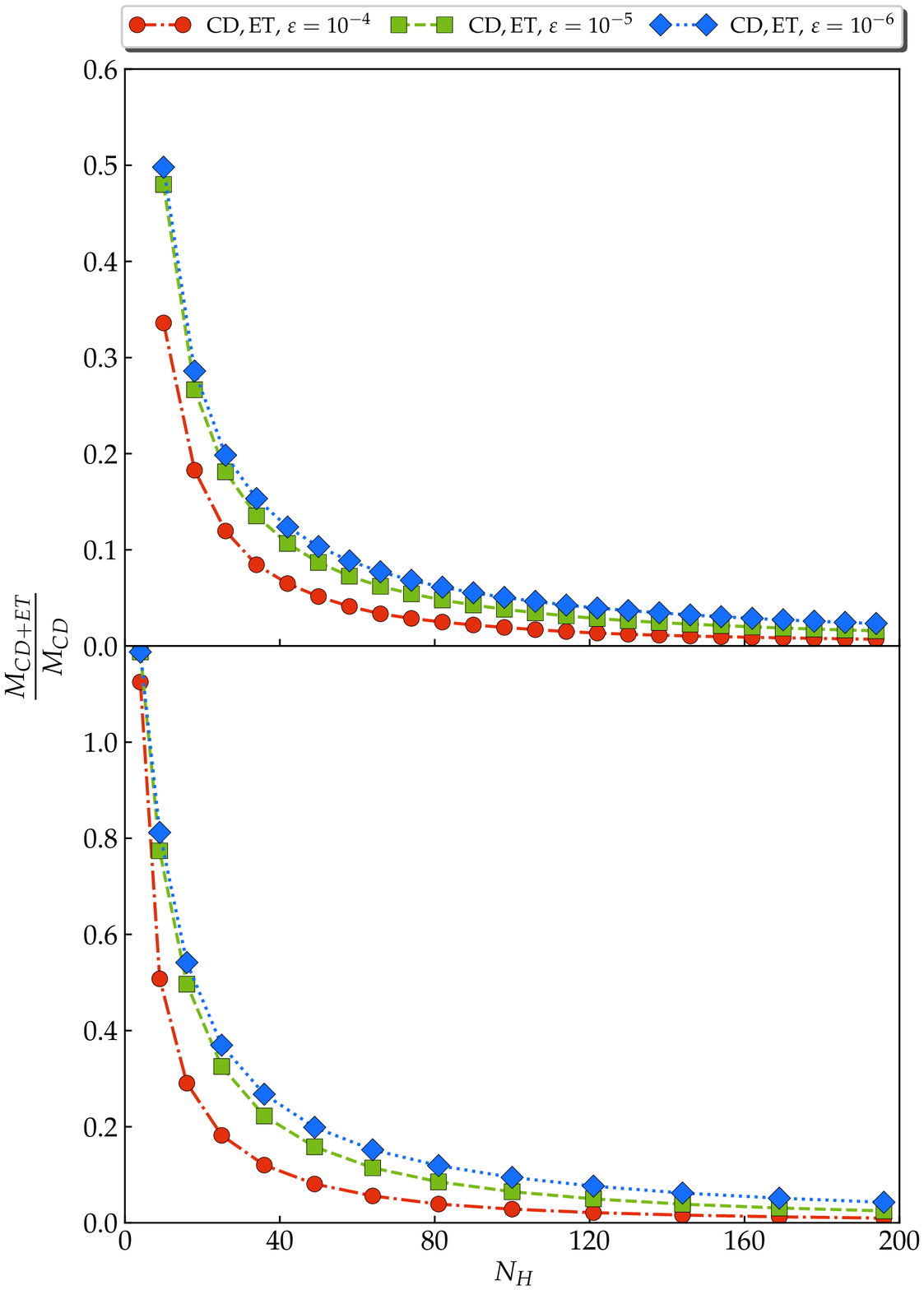}
\caption{
(color online) Ratio between the memory required for local energy precomputing 
for AFQMC with CD ($M_{CD}$) and AFQMC with CD+ET $(M_{CD+ET})$, 
as a function of the number $N_H$ of hydrogen atoms for H chains (top) and square grids (bottom) 
at the representative bondlength $R = 1.8$ a.u. at STO-6G level of theory.
Three different truncation thresholds, $\varepsilon = 10^{-4}, 10^{-5}, 10^{-6}$ a.u. are explored
(red circles, green squares, blue diamonds).
\label{fig:memory_reduction}
}
\end{figure}

The memory reduction from the low-rank factorization is shown
in Figure \ref{fig:memory_reduction}, where the ratio between the size 
of the tensors $\bar{V}$ and $A$, $B$ is shown for hydrogen chains and grids.
As seen, for a large system, the size of $A$, $B$ is only $\simeq 5 \%$ of that of $\bar{V}$.

\REVISION{
Note that eigenvalue truncation should not be performed
for the mean-field energy, because it effectively truncates the Coulomb interaction leading to
an incorrect treatment of the classical electrostatics of the electron distribution (overall charged without the nuclei) from truncating charge-charge terms.
Consequently, we compute the mean-field energy with and without CD+ET, and add the corresponding correction to the local energy computed from Eq.~\ref{eq:ftensor},
\begin{equation}
  \label{eq:correctafqmc}
E_{AFQMC} = \left( E_{AFQMC}^{CD+ET} - E_{RHF}^{CD+ET} \right) + E_{RHF} \,\, .
\end{equation}

Since we achieve a reduction in cost because the truncated low-rank factorization implements
a form of integral screening, we briefly compare the low-rank factorization to
directly screening the local energy evaluation in Eq.~\eqref{eq:localenergy2}. A direct screening
of the sum in Eq.~\eqref{eq:localenergy2} will give $\mathcal{O}(N^2)$ cost for moderate sized
and systems, and asymptotically $\mathcal{O}(N)$ cost (although formation of the Green's function and
other operations still require $\mathcal{O}(N^3)$ cost).
However, unless one evaluates the ERI
on the fly, this requires $\mathcal{O}(N^4)$ storage, and, for larger basis sets, there will be
a very large number of significant ERIs, as one does not achieve the basis compression
afforded by CD or DF. Thus, 
except for large systems in small basis sets, we expect the CD+ET approach to be superior to simple integral screening.

A hybrid strategy would be to use the sparsity of the Cholesky vectors directly (e.g. implement the construction
of all intermediates in the local energy using sparse matrix multiplication). Compared to low-rank
factorization, sparse matrix multiplication often incurs overhead for smaller problem sizes. However, a detailed
comparison between the direct use of sparsity in the Cholesky vectors versus the nested matrix factorization is an interesting question
to explore in the future.}

\section{Results}
\label{sec:res}

We now apply the formalism outlined in Section \ref{sec:thc} to several test systems, 
including both molecules and crystalline solids.
In each case we compare the local energy evaluation time $T_{Eloc}$ from conventional 
AFQMC and AFQMC with CD+ET, assess the accuracy of the ET procedure, and investigate 
the scaling with system size.
Timing calculations were performed on a cluster with nodes having 2 CPUs with 14 cores each (Intel E5-2680, 2.4 GHz).

\subsection{Networks of H atoms}

\begin{figure}[ht!]
\centering
\includegraphics[width=0.4\textwidth]{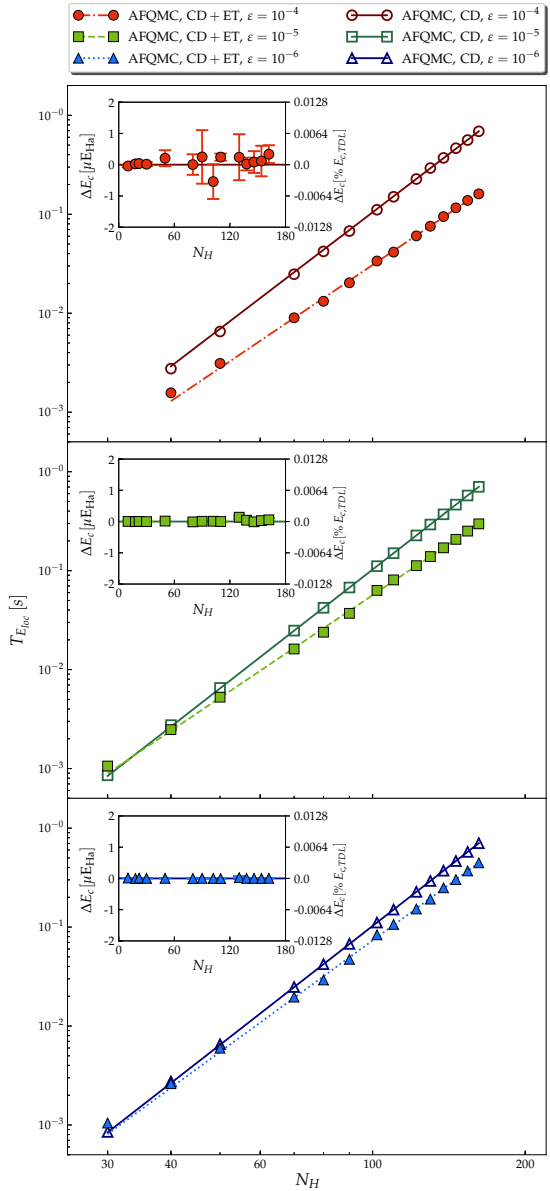}
\caption{
(color online) Main figures: Log-log plot of the local energy evaluation time $T_{E_{loc}}$ as a function of the number $N_H$ of
hydrogen atoms for H chains at the representative bondlength $R = 1.8$ a.u. at the STO-6G level of theory,
from AFQMC with CD (empty markers) and AFQMC with CD+ET (filled markers).
Truncation thresholds, $\varepsilon = 10^{-4}, 10^{-5}, 10^{-6}$ a.u. (top to bottom) are explored.
Solid, dashed lines are the result of fit of AFQMC with CD, CD+ET to \eqref{eq:timing}.
Insets: average difference in the correlation part of the local energy, per atom, between AFQMC with CD
and CD+ET.
}
\label{fig:h1}
\end{figure}

We first consider the test case of hydrogen (H)
chains \cite{Hachmann2006,Motta_PRX_2017}, at a representative bondlength $R=1.8 \, \mathrm{a_B}$, 
using the minimal STO-6G basis and RHF trial wavefunction.
We use identical thresholds for CD and eigenvalue truncation, $\varepsilon_{CD} = 
\varepsilon_{ET} = 10^{-4}$, $10^{-5}$, $10^{-6}$ a.u.
The local energy evaluation time (per walker, averaged over many walkers) using the conventional AFQMC formula \REVISION{with
Cholesky decomposition}, Eq. \eqref{eq:convafqmc}, and 
CD+ET-based AFQMC (CD+ET-AFQMC) of Section \ref{sec:thc}, is shown in Figure \ref{fig:h1}.
\REVISION{The reported times reflect the cumulative impact of floating-point and fundamental 
memory operations (e.g. allocations of arrays). The overhead due to memory operations, which
we estimate of the order of $\sim 1$ ms, becomes increasingly less important as the size of
the studied systems increases.
Local energy calculations times are reproduced well by the formulae
\begin{equation}
\begin{split}
&\REVISION{T_{CD}} \simeq t_0 \, N^\alpha \quad,\quad T_{CD+ET} \simeq t_0 \, N^\beta \quad .
\end{split}
\label{eq:timing}
\end{equation}
We observe  exponents $\alpha = 3.91(2)$, $3.99(1)$, $3.99(1)$ and $\beta = 3.46(3)$, $3.54(2)$, $3.74(3)$ 
for $\varepsilon = 10^{-4}$, $10^{-5}$, $10^{-6}$ a.u. respectively.
The untruncated local energy calculation displays the anticipated quartic scaling, while the looser CD+ET truncation 
thresholds reach sub-quartic scaling for these system sizes. Given the relationship between ET and integral screening, 
it is unsurprising that saturation of $\langle \rho_\gamma \rangle$ (responsible for cubic scaling) is not reached for
the tightest threshold. Nonetheless, the local energy evaluation time is still reduced relative to using only CD.}
The prefactors in the two functions determine the number $N_H^*$ of H atoms required for the two curves to cross. 
We find that $N^*_H \simeq 25, 35,40$ for the three thresholds we have considered.

In the insets, we compute the difference $\Delta E_c$ between the correlation energies per atom
from AFQMC and CD+ET, as function of the number of H atoms, using the estimator
\begin{equation}
\Delta E_c = \frac{1}{N_w} \sum_{w} [E_{loc,c}(\Phi_w) - E^\prime_{loc,c}(\Phi_w)] \quad ,
\end{equation}
where $E_{loc,c}(\Phi) = E_{loc}(\Phi) - E_{HF}$ is defined in terms of the standard local
energy functional \eqref{eq:convafqmc}, but using integrals reconstructed from the CD vectors,
while $E^\prime_{loc,c}(\Phi) = E^\prime_{loc}(\Phi) - E^\prime_{HF}$
is formulated in terms of the CD+ET expression, Section \eqref{sec:thc}, for the local energy.
In Figure \ref{fig:h1}, $\Delta E_c$ is evaluated on 6 independently generated populations 
of walkers equilibrated for $\beta=2 E_{Ha}^{-1}$.
Using all thresholds, the energies per atom agree to within $0.02 \%$ of the 
total correlation energy extrapolated to the thermodynamic limit (TDL), confirming the good accuracy
of the CD+ET decomposition for conservative choices of the threshold $\varepsilon$.

\begin{figure}[ht!]
\centering
\includegraphics[width=0.49\textwidth]{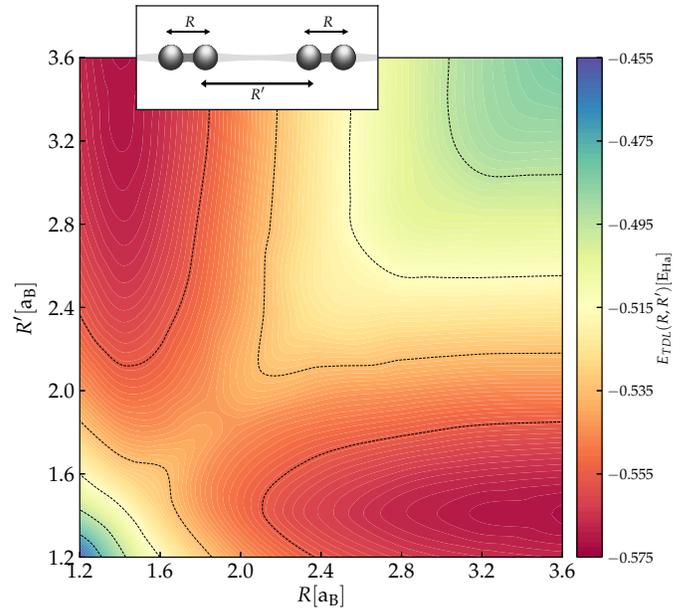}
\caption{(color online) Energy per atom of the H chain at the STO-6G level of theory, as
a function of the intra-bond and inter-bond lengths $R,R^\prime$.
Results are obtained for $R,R^\prime = 1.2,1.4,1.6,1.8,2.0,2.4,2.8,3.2,3.6$ a.u., and 
the potential energy surface is produced via cubic spline interpolation.
}
\label{fig:surf}
\end{figure}

\REVISION{As a simple application, 
we next study the asymmetric dissociation of the infinite H chain using the STO-6G basis in Figure \ref{fig:surf}.
Note that, as seen in Figure \ref{fig:h1}, the cost of CD+ET calculations is reduced
by a factor of roughly 10 for $N_H=100$, so the full study of the asymmetric dissociation can be carried out at
appreciably reduced computational cost, comparable with the cost of studying the symmetric dissociation only.}

More specifically, we compute the potential energy surface of a 
network of H atoms at positions ${\bf{R}}_{k,\pm} = (0,0,z_{k,\pm})$ with 
$z_{k,\pm} = \pm \frac{R}{2} + k (R^\prime+R)$, $k=0 \dots \frac{N}{2}-1$, for a total number
of atoms between $N_H=10$ and $N_H=100$, as function of the intra-bond and inter-bond 
lengths $R$, $R^\prime$. We use the UHF Slater determinant \REVISION{as a trial wavefunction}. For all $R,R^\prime$ in a mesh of points
between $1.2$ and $3.6$ $\mathrm{a_B}$, we extrapolate the energy per atom $E(R,R^\prime,N)$
to the TDL using standard procedures \cite{Motta_PRX_2017}, and compute 
correlation energies using AFQMC with CD+ET and the truncation threshold
$\varepsilon = 10^{-5}$ au.
The extrapolated potential energy surface $E(R,R^\prime) = \lim_{N \to \infty} E(R,R^\prime,N)$
is shown in Figure \ref{fig:surf}, and values for $R^\prime=1.6$, $2.4$, $3.2 \, \mathrm{a_B}$ are
given in Table \ref{tab:surf}.

The diagonal of Figure \ref{fig:surf} corresponds to the symmetric dissociation of the 
chain, $R=R^\prime$ \cite{Motta_PRX_2017}, the minimum energy being reached 
at the saddle point $R = R^\prime  \simeq 1.83 \, \mathrm{a_B}$.
For large $R, R^\prime$ the potential energy surface increases towards the energy
$E_H = 0.471 \, \mathrm{E_{Ha}}$ of a single H atom in the STO-6G basis, and the global minimum 
of the energy is reached for $R^\prime \to \infty$, $R \simeq 1.4 \, \mathrm{a_B}$, 
corresponding to a collection of uncoupled H$_2$ molecules, with energy 
$E_{\mathrm{H_2}} = -0.573 \, \mathrm{E_{Ha}}$. This illustrates
the well-known Peierls instability of equally spaced atomic chains under lattice distorsions.

\begin{table}[t!]
\begin{tabular}{cccc}
\hline\hline
$R^\prime$ & $E(R=1.6,R^\prime)$ & $E(R=2.4,R^\prime)$ & $E(R=3.2,R^\prime)$ \\
\hline
1.2 & -0.51517(6) & -0.55578(9) & -0.5652(1) \\
1.4 & -0.52857(9) & -0.5619(2) & -0.5704(1) \\
1.6 & -0.53362(6) & -0.5582(2) & -0.5660(3) \\
1.8 & -0.54288(8) & -0.5507(2) & -0.5569(4) \\
2.0 & -0.5498(1)  & -0.5411(1) & -0.5454(3) \\
2.4 & -0.5582(2)  & -0.5233(1) & -0.5223(1) \\
2.8 & -0.5634(2)  &  -0.5219(1) & -0.5037(1) \\
3.2 & -0.5660(2)  & -0.5223(1) & -0.4915(1) \\
3.6 & -0.5672(2)  & -0.5228(1) & -0.4902(1) \\
\hline\hline
\end{tabular}
\caption{Energy per atom of the H chain at the STO-6G level of theory, extrapolated to the
thermodynamic limit, as a function of the inter-bond length $R^\prime$ for $R=1.6,2.4,3.2$ a.u. (left to right).
} \label{tab:surf}
\end{table}

\begin{figure}[ht!]
\centering
\includegraphics[width=0.4\textwidth]{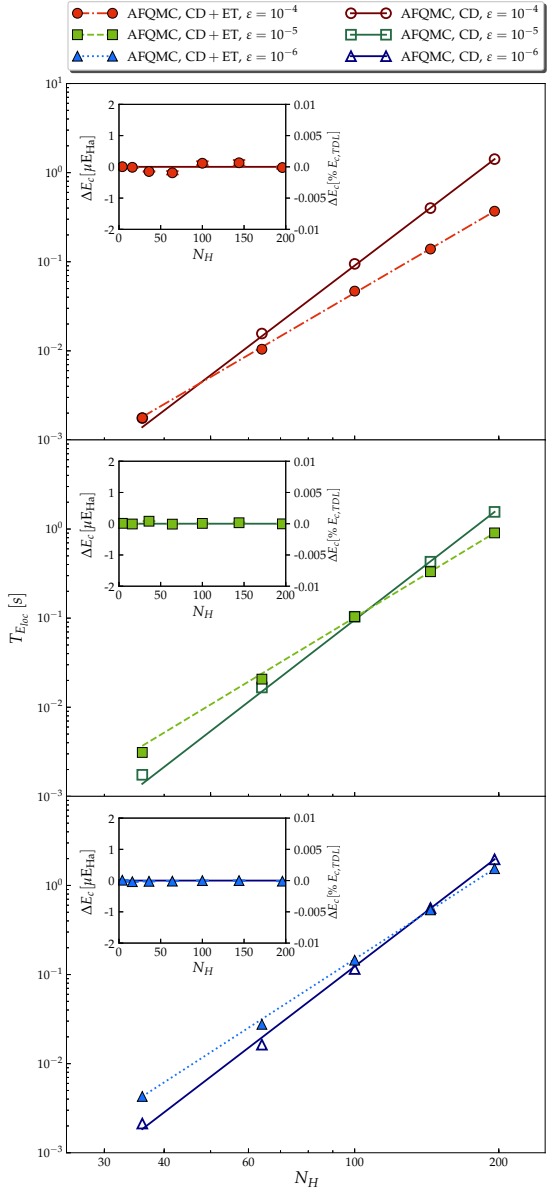}
\caption{(color online) Main figures: local energy evaluation time as a function of the number $N_H$ of
hydrogen atoms, for H square grids at the representative bondlength $R = 1.8$ a.u. at
the STO-6G level of theory, from AFQMC with CD (empty markers)
and AFQMC with CD and ET (filled markers).
Crossover between the two strategies is seen for $N_H \simeq 50, 100, 150$ for
increasingly small threshold.
Insets: average difference in the correlation part of the local energy, per atom, between
state-of-the-art AFQMC and AFQMC with eigenvalue truncation.
}
\label{fig:h2}
\end{figure}

We continue our assessment of accuracy and performance by studying, in Figure
\ref{fig:h2}, two-dimensional square grids of H atoms, where the H atoms 
occupy positions ${\bf{R}}_{ij} = (0,i R,j R)$, $i,j = 0 \dots n-1$. Here $n$ is
related to the number $N_H$ of atoms in the grid as $N_H = n^2$, and
we work at the representative bondlength $R=1.8 \, \mathrm{a_B}$. 

The trends seen for H chains are confirmed: the standard and CD+ET-based 
local energy calculation times are well described by \eqref{eq:timing} with 
exponents \REVISION{$\alpha = 4.09(2)$, $4.14(3)$, $4.11(3)$ and 
$\beta = 3.14(2)$, $3.25(1)$, $3.47(2)$ for $\varepsilon = 10^{-4}, 10^{-5}, 10^{-6}$ 
  au respectively. (Note that the $\beta$ exponents here are slightly lower than in 1D which may seem surprising,
  but the model analysis shows that $\langle \rho_\gamma\rangle$ as a function of system size in 1D and 2D
  can actually crossover before saturation, with the 1D curve growing more steeply,
  but saturating at smaller system size, than the 2D curve).}
Crossover between the two approaches is seen for $N_H^* \simeq 50, 120, 170$ for 
increasingly small threshold.
The discrepancy $\Delta E_c$ between correlation energies based on \REVISION{AFQMC 
with CD} and AFQMC with CD+ET is consistently below $0.01 \%$ of the correlation energy 
per atom extrapolated to the TDL, further confirming the accuracy of the truncation scheme.

\subsection{Water clusters}

\begin{figure}[ht!]
\centering
\includegraphics[width=0.45\textwidth]{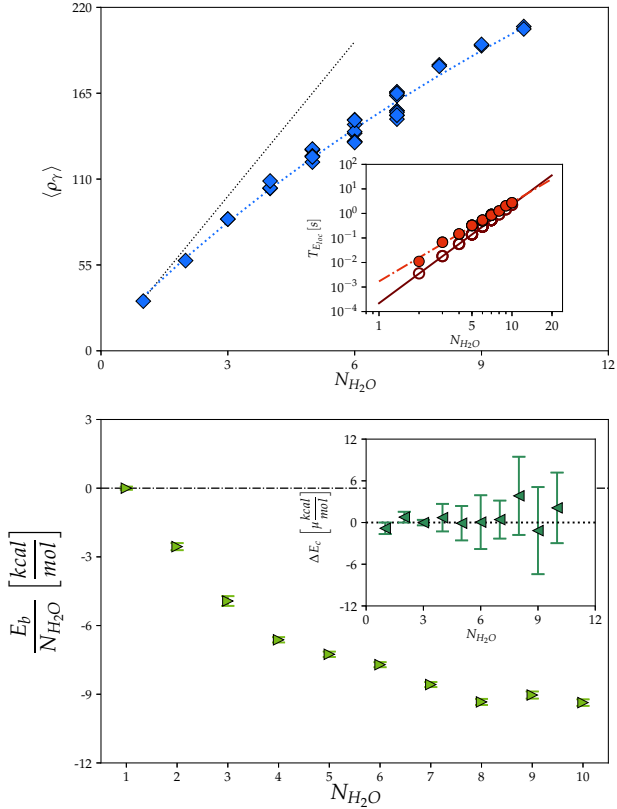}
\caption{(color online) Top: average number $\langle \rho_\gamma\rangle$ of retained
eigenvalues as function of the number of H$_2$O molecules in small water clusters
\cite{Temelso2011}, using $\varepsilon = 10^{-4}$ au.
Inset: local energy calculation time from AFQMC with CD and CD+ET (empty, filled symbols).
Solid, dot-dashed lines indicate fit to \eqref{eq:timing}.
Bottom: AFQMC binding energy per monomer, for the most stable water clusters with given number $N_{\mathrm{H_2O}}$ of monomers.
Inset: difference $\Delta E_c$ between correlation energy per monomer from AFQMC with CD and with CD+ET, for all clusters.
}
\label{fig:h2o}
\end{figure}

To test larger basis sets and heavier elements, in Figure \ref{fig:h2o}, we
investigate $38$ water clusters (motivated by studies of water clusters in the terrestrial atmosphere) 
containing 2-10 water molecules \cite{Temelso2011}, 
using the heavy-augmented cc-pVDZ basis (aug-cc-pVDZ for O, cc-pVDZ for H),
\REVISION{a truncation threshold $\varepsilon = 10^{-4}$ a.u. and a} RHF trial wavefunction.
Also in this case, the average number of retained eigenvalues $\langle \rho_\gamma \rangle$
grows \REVISION{sub-linearly} with the size of the system, as measured by the number
of H$_2$O molecules, leading to a \REVISION{sub-quartic scaling local energy evaluation 
(upper panel)}.
The dependence of the local energy calculation time on the number of water molecules,
shown in the inset of the upper panel, is again well represented by Eq. \eqref{eq:timing} 
with $\alpha=4.01(1)$ and \REVISION{$\beta = 3.21(3)$}, so that crossover between 
conventional \REVISION{AFQMC with CD} and CD+ET local energy calculation times
is seen at $N_{\mathrm{H_2O}} \simeq 13$.

In the inset of the lower panel, we show the difference $\Delta E_c$ between the 
correlation energies per atom from AFQMC with CD integrals and CD+ET, as a function of the 
number of monomers. $\Delta E_c$ is evaluated on 6 independently generated populations 
of walkers equilibrated for $\beta=2 E_{Ha}^{-1}$ and, for a given cluster size
 $\Delta E_c$ is averaged over all cluster structures with the same number of monomers.
For example, for $N=5$, $\Delta E_c$ is averaged over the 6 water pentamers labelled
CYC, CAA, CAB, CAC, FRA, FRB, FRC in \cite{Temelso2011}.
The error is only a few $\mu$ kcal/mol.

The binding energy per water molecule for the most stable clusters,
labelled 2Cs, 3UUD, 4S4, 5CYC, 6PR, 7PR1, 8D2d, 9D2dDD, 10PP1 
in \cite{Temelso2011}, is shown in the lower panel of Figure \ref{fig:h2o}.
As seen, the binding energy per molecule decreases almost monotonically with
the number of monomers in the cluster, reaching $E_b / N_{\mathrm{H_2O}} \simeq -9$ kcal/mol for 
$N_{\mathrm{H_2O}} \geq 8$.

\begin{figure}[h!]
\centering
\includegraphics[width=0.45\textwidth]{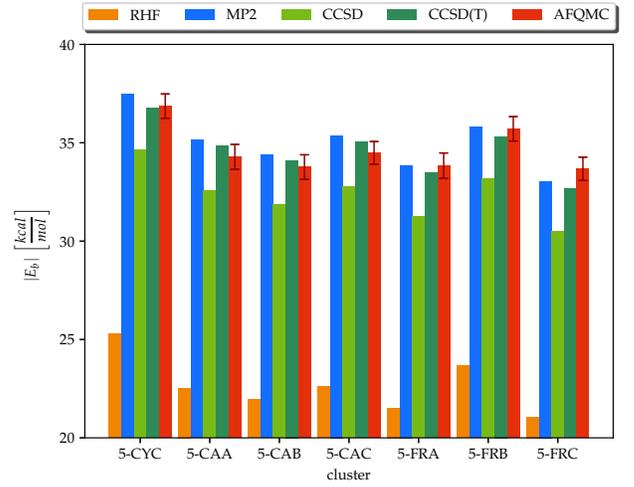}
\caption{
Binding energy for water pentamers by RHF, MP2, CCSD, CCSD(T) and AFQMC(CD+ET), in kcal/mol,
using the heavy-augmented cc-pVDZ basis.
} \label{tab:clus}
\end{figure}

Numerical data supplied in Table \ref{tab:clus_stable} provide a comparison with RHF,
MP2, CCSD and CCSD(T). Energies from these methodologies are computed 
without performing any truncation on the Hamiltonian, while AFQMC energies are 
estimated using Eq. \eqref{eq:correctafqmc}.
As seen, correlated methods are in relatively good agreement with each other. 
AFQMC is in good agreement with CCSD(T), with an average deviation
of $\Delta = -0.59(29)$ kcal/mol.
Data for the different water pentamers are showed in Figure \ref{tab:clus}.
Binding energies from correlated methods are in good agreement with each other and display the same trends.
The  average deviation between CCSD(T) and AFQMC is $\Delta = 0.06(61)$ kcal/mol.

\begin{table}
\centering
\begin{tabular}{lccccc}
\hline\hline
cluster &
$E_{b,RHF}$ &
$E_{b,MP2}$ &
$E_{b,CCSD}$ &
$E_{b,CCSD(T)}$  &
$E_{b,AFQMC}$  \\
\hline
2Cs      &  -3.815 & -5.217  & -4.912  & -5.179 &   -5.11(31) \\
3UUD     & -10.521 & -15.833 & -14.670 & -15.619 & -14.78(64) \\
4S4      & -19.001 & -28.358 & -26.210 & -27.865 & -26.49(46) \\
5CYC     & -25.297 & -37.482 & -34.627 & -36.776 & -36.27(59) \\
6PR      & -29.917 & -47.246 & -43.727 & -46.823 & -46.25(66) \\
7PR1     & -37.486 & -59.149 & -54.619 & -58.470 & -60.04(77) \\
8D2s     & -46.650 & -74.924 & -69.023 & -74.044 & -74.7(1.1) \\
9D2dDD   & -53.395 & -84.816 & -78.164 & -83.739 & -81.3(1.4) \\
10PP1    & -60.449 & -96.615 & -89.029 & -95.453 & -93.7(1.4) \\
\hline\hline
\end{tabular}
\caption{Binding energy for the most stable water clusters reported in \cite{Temelso2011},
by RHF, MP2, CCSD, CCSD(T) and AFQMC(CD+ET), in kcal/mol,
using the heavy-augmented cc-pVDZ basis.}
\label{tab:clus_stable}
\end{table}

\subsection{Two-dimensional hexagonal boron nitride}

We now consider a crystalline solid, 2D hexagonal boron nitride (BN).
To perform these calculations we used an underlying single-particle basis of crystalline 
Gaussian-based atomic orbitals, which are translational-symmetry-adapted linear 
combinations of Gaussian atomic orbitals \cite{Berkelbach_JCTC_2017}.
Core electrons were replaced with separable norm-conserving \REVISION{GTH-LDA} 
pseudopotentials
\cite{Goedecker_PRB_1996, Hartwigsen_PRB_1998}, removing sharp nuclear
densities.
Matrix elements for the Hamiltonian of the system were computed with the PySCF
\cite{Sun_WIRES_2018} package, using the GTH-DZV Gaussian basis set 
\cite{Lundqvist_PKM_1967}. The RHF state was used as trial wavefunction.

\begin{figure}[ht!]
\centering
\includegraphics[width=0.42\textwidth]{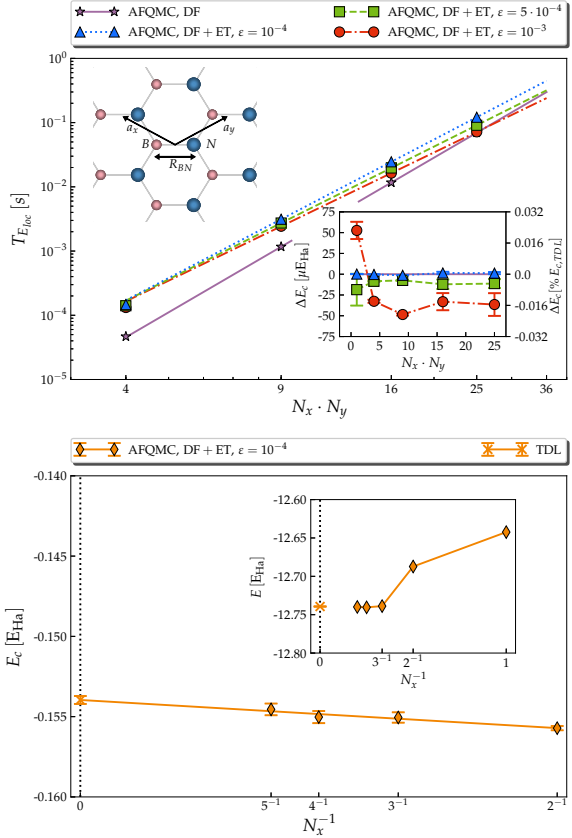}
\caption{
(color online) Top: Local energy evaluation time $T_{E_{loc}}$ from AFQMC with DF approximation
for the ERI (purple stars), and AFQMC with DF+ET (red circles, green squares, blue triangles for 
$\varepsilon = 10^{-3}$, $5 \cdot 10^{-4}$, $10^{-4}$ a.u. respectively), for 2D hexagonal BN at $R_{BN} = 2.5$\AA, 
as function of supercell size. Inset: difference in the correlation part of the local energy per unit cell.
Bottom: extrapolation to the thermodynamic of AFQMC correlation (main plot) and total (inset) energy (orange
diamonds) per unit cell. Extrapolations to the \REVISION{thermodynamic limit} 
are indicated by orange crosses.
}
\label{fig:bn}
\end{figure}

Size effects were removed studying increasingly large supercells at the $\Gamma$ point.
Supercells were obtained repeating the primitive, two-atom cell $N_x = N_y = 1, \dots ,5$ 
times along directions $a_x$, $a_y$ sketched in Figure \ref{fig:bn}, and we operated at the
representative bondlength $R_{BN} = 2.5${\AA} to illustrate the effects of the eigenvalue
truncation on top of the DF approximation.
The ERI was obtained using the Gaussian DF approximation~\cite{GDFPW},
\REVISION{
\begin{equation}
(pr|qs) = \sum_L (pr|L) \Omega_{LM}^{-1} (qs|M) = \sum_\gamma D^\gamma_{pr} D^\gamma_{qs}
\end{equation}
with $D^\gamma_{pr} = \sum_L (pr|L) \Omega^{-1/2}_{L\gamma}$, and eigenvalue 
truncation was performed on the DF operators $D^\gamma_{pr}$} with truncation
thresholds $\varepsilon = 10^{-4}$, $5 \cdot 10^{-4}$, $10^{-3}$ a.u.

In the upper panel of Figure \ref{fig:bn} we illustrate the local energy evaluation time
from AFQMC with DF and DF+ET as a function of supercell size $N_x \cdot N_y$.
Crossover is seen for $N_x \simeq 5,6$ for increasingly small thresholds.
For a widegap semiconductor like BN, as discussed below, supercells of this size are
sufficient to converge mean-field and correlation energies to the thermodynamic limit. 
We thus expect the DF+ET approach to be even more beneficial for materials with smaller 
or vanishing gap (e.g. metals), that require even larger supercells or Brillouin zone
meshes to reliably converge energies to the TDL.
\REVISION{In the lower panel of Figure \ref{fig:bn}, we extrapolate the AFQMC correlation
energy to the TDL using the power-law Ansatz $E_c(N_x) = \alpha + \beta N_x^{-1}$
\cite{Berkelbach_JCTC_2017}. We add to the extrapolated AFQMC correlation energy $\alpha$ 
the extrapolated RHF energy, obtained following the procedure in \cite{Berkelbach_JCTC_2017}.
}
The extrapolated total energy is shown in the inset of the lower panel of Figure \ref{fig:bn}.

In the inset of the upper panel, we illustrate the difference $\Delta E_c$ between the 
correlation energy per cell from AFQMC with DF and DF+ET, estimated on 3 populations
of walkers equilibrated for $\beta = 4 \, \mathrm{E^{-1}_{Ha}}$.
As naturally expected, $\Delta E_c$ increases monotonically with the truncation threshold, though 
remaining consistently below $0.03 \%$ of the AFQMC correlation energy extrapolated 
to the TDL.

\section{Conclusions}
\label{sec:conc}

In the present work we have shown that, through a simple and efficient low-rank factorization of the ERI, 
it is possible to reduce the asymptotic complexity of AFQMC calculations for electronic structure problems 
in Gaussian bases \REVISION{from its conventional quartic scaling with system size.
While the asymptotic cubic scaling is attained only for large systems, at the intermediate sizes 
studied we nevertheless observed sub-quartic scaling accompanied by significant memory savings and 
high accuracy.}
This reduction arises from exposing the sparsity of ERI in Gaussian bases through a nested matrix 
diagonalization. This approach will be useful in studies of larger molecules, and of crystalline 
solids requiring extrapolations to the thermodynamic limit. \REVISION{We also find that the memory requirements using this approach 
are significantly reduced from the conventional AFQMC algorithm using Cholesky decomposition alone.}
The algorithmic advances may also be used in conjunction with parallel efforts to accelerate AFQMC 
through improved hardware implementations~\cite{Shee_JCTC_2018}.
\REVISION{While more work is necessary to
   establish the relative benefits in AFQMC of exposing sparsity through low rank, as in the current work, versus
   the direct utilization of sparse operations, we expect such combinations to greatly advance
   the practical possibilities for AFQMC calculations
 on large systems.}

\section{Acknowledgments}

 M. M. and G. K. C. were supported by the US NSF (Grant No. 1665333).
 S. Z. acknowledges support from DOE (Grant No. DE-SC0001303).
 Additional
 software developments for the AFQMC periodic calculations implemented in PySCF were supported by
 US NSF (Grant No. 1657286).
Computations were carried out on facilities supported by the National Energy Research Scientific Computing Center (NERSC),
on facilities supported by the Scientific Computing Core at the Flatiron Institute, a division of the Simons Foundation,
on the Pauling cluster at the California Institute of Technology, 
and on the Storm and SciClone Clusters at the College of William and Mary.
M. M. acknowledges Narbe Mardirossian, Yuliya Gordiyenko and Qiming Sun for useful 
discussion about electronic structure calculations for H$_2$O clusters and BN.

\appendix 

\section{Relationship with plane-wave formulations}
\label{sec:pw}

Many calculations in solid-state systems are often performed using a plane wave basis. 
AFQMC simulations using this computational basis~\cite{Zhang_PRL90_2003,Suewattana_PRB75_2007} have an $\mathcal{O}(N^3)$ scaling.
We here briefly outline the relationship between the cubic scaling achieved in the plane-wave basis and that achieved
using the factorization techniques in this paper. 

In the plane-wave basis, the Hamiltonian with pseudopotentials takes the form
\begin{equation}
\begin{split}
\label{eq:hampw}
\hat{H} &= H_0 + \sum_{ \vett{G} \vett{G}^\prime} t_{\vett{G}\vett{G}^\prime} \hat{E}_{\vett{G}\vett{G}^\prime} + \frac{1}{2}   \, \sum_{ \vett{G}  \vett{G}^\prime \vett{q} } V_\vett{q} \hat{E}_{\vett{G}+\vett{q} \vett{G}} \hat{E}^\dag_{\vett{G}^\prime\vett{G}^\prime+\vett{q}} 
\end{split}
\end{equation}
where $\vett{G}$ is a wave-vector in the reciprocal lattice, corresponding
to the plane-wave state $\langle \mathbf{r}|\mathbf{G}\rangle = \frac{ e^{i \vett{G} \cdot \vett{r} } }{ \sqrt{\Omega} }$ where $\Omega$
is the computational cell volume.
The vectors $\vett{q}$ are the transfer momenta, and, due to momentum conservation, their number is proportional to
number of plane-waves $N$. Thus Eq.~\eqref{eq:hampw} is a low-rank factorization of the integrals (with $\vett{q}$ playing
the role of $\gamma$) but it is not
a Cholesky factorization, because the analogous quantity
\begin{align}
L^{\vett{q}}_{\vett{G} \vett{G}^\prime} 
= \sqrt{V_\vett{q}} \delta_{\vett{G},\vett{G}^\prime+\vett{q}} \label{eq:periodicL}
\end{align}
is not a lower triangular matrix for each $\vett{q}$.

The local energy formula can be written analogously to \eqref{eq:energy_from_f},
\begin{equation}
\mathcal{E}_{loc,2}(\Phi) = 
 \sum_{ij\vett{q}} f^\vett{q}_{ii} f^{-\vett{q}}_{jj} 
- \, f^{\vett{q}}_{ij} f^{-\vett{q}}_{ji} \, ,
\end{equation}
with $f^\vett{q}_{ij}$ defined formally as
\begin{equation}
  \label{eq:periodicf}
  f^\vett{q}_{ij} = \sum_{\vett{G} \vett{G}^\prime} L^{\vett{q}}_{\vett{G} \vett{G}^\prime} 
 {\Phi_T}_{i \vett{G}} \Theta_{\vett{G}^\prime j} 
  \end{equation}
Unlike in the case of the atomic orbital basis, the $L^\vett{q}$ matrices contain $N$, 
rather than $\mathcal{O}(\log N)$, elements, and do not display the same low-rank structure. However, $L^\vett{q}$ encodes a periodic delta function,
which means that \eqref{eq:periodicL} is a convolution,
 \begin{equation}
 f^\vett{q}_{ij} = \sum_{\vett{G}} \Phi^{\sigma}_{i \vett{G}} \Theta^\sigma_{\vett{G}+\vett{q} j}.
 \end{equation}
Consequently, using the fast Fourier transform, computing $ f^\vett{q}_{ij}$ requires only $\mathcal{O}(O^2N \log N)\sim \tilde{\mathcal{O}}(N^3)$ time, and
 the local energy evaluation can be computed in soft cubic time.


%

\end{document}